# Estimation of gridded population and GDP scenarios with spatially explicit statistical downscaling

**Daisuke Murakami\* and Yoshiki Yamagata**

Center for Global Environmental Research, National Institute for Environmental Studies, 16-2, Onogawa, Tsukuba, Ibaraki 305-8506, Japan.

\* Corresponding author. E-mail: murakami.daisuke@nies.go.jp

**Abstract**: This study downscales the population and gross domestic product (GDP) scenarios given under Shared Socioeconomic Pathways (SSPs) into 0.5-degree grids. Our downscale approach has the following features: (i) it explicitly considers spatial and socioeconomic interactions among cities; (ii) it utilizes auxiliary variables, including, road network and land cover; (iii) it endogenously estimates influence from each factor by a model ensemble approach; (iv) it allows us controlling urban shrinkage/dispersion depending on SSPs. It is confirmed that our downscaling results are consistent with scenario assumptions (e.g., concentration in SSP1 and dispersion in SSP3). Besides, while existing grid-level scenario tends to have overly-smoothed population distributions in non-urban areas, ours does not suffer from the problem, and captures difference in urban and non-urban areas in a more reasonable manner.

1. Introduction

Socioeconomic scenarios are needed to project carbon dioxide ($CO_2$) emissions, disaster risks, and other factors affecting sustainability from a long-term perspective. The Intergovernmental Panel on Climate Change (IPCC) published Spared Socioeconomic Pathways (SSPs; O'Neill et al., 2014; 2017) that describe future socioeconomic conditions under various scenarios, including SSP1-3. SSP1 makes relatively good progress toward sustainability under an open and globalized world. SSP2 is a middle-of-the-road scenario assuming that the typical trends in the last decades will continue, and in SSP3, the world is closed and fragmented into regions, and it fails to achieve sustainability.

While the SSPs are devised in terms of country scenarios, finer scenarios (e.g., scenarios in terms of 0.5-degree grids) are required to analyse regional/city-level sustainability and resiliency. A number of studies downscale country-level socioeconomic scenarios into finer spatial units (e.g., Bengtsson et al., 2006; Asadoorian, 2007; Grübler et al., 2007; van Vuuren et al., 2007; Gaffin et al. 2011; Hachadoorian et al., 2011; Jones and O'Neill, 2013; Nam and Reilly, 2013; McKee et al., 2015; Yamagata et al., 2015; Jones and O'Neill, 2016).

Yet, these studies have several limitations. First, (i) they do not consider interactions among cities. It is likely that spatial interactions, that is, interactions depending on geographical distance, are significant locally, while economic interactions are significant globally. These local and global interactions among cities must be considered in addition to spill-over from cities to their neighbours, as done by Jones and O'Neill (2013).

Second, (ii) many of previous studies do not utilize auxiliary variables (e.g., landuse, road network, location of airports), which seem useful for the consideration of urban form and functions, but rather, simply extrapolate past trend using logistic-of-growth model (e.g., Gaffin et al., 2004), share-of-growth model (e.g., Yamagata et al., 2015), gravity-type model (e.g., Grübler et al., 2007; Jones and O'Neill, 2016), and so on. McKee et al.'s (2015) study is an exception, as it considers landuse data, road network data, and so on. Nevertheless, their target area is limited to the USA. Also, they determine weights exogenously for each auxiliary variable. It is desirable to estimate the importance of each auxiliary variable endogenously.

The objective of this study is downscaling the country-level SSP1-3 scenarios into 0.5-degree grids while overcoming the two limitations. Specifically, our downscale approach estimates (i) intensity of interactions among cities and (ii) importance of auxiliary variables, from data. Although Jones and O'Neill (2016) already published gridded SSP population scenarios (https://www2.cgd.ucar.edu/sections/tss/iam/ssp-projections), they apply a simple approach ignoring auxiliary variables. Our study considering (i) and (ii) would be beneficial to develop a more sophisticated gridded scenarios.

**2. Overview of our downscaling approach**

This study downscales the urban population, non-urban population, and gross domestic product (GDP; Purchasing power parity (PPP), Billion USD in 2005 year rate) by country under SSP1-3 (Source: SSP Database: https://tntcat.iiasa.ac.at/SspDb/dsd?Action=htmlpage&page=about) into 0.5-degree grids. Urban and non-urban populations are obtained by dividing country populations using share of urban populations projected by Jiang and O'Neill (2015). The target years are from 2010 to 2100 by five years.

Figure 1 summarizes our downscaling approach. Urban populations by country are downscaled into cities based on a city growth model. The estimated city populations are used to estimate urbanisation potential that is used to project urban expansion. The city populations are further downscaled into 0.5-degree grids considering projected urban expansion and auxiliary variables summarize in Table 1. Non-urban populations are downscaled into 0.5-degree grids considering urban expansion and the auxiliary variables. GDP is also downscaled considering urban expansion, the auxiliary variables, and downscaled populations.

Hereafter, Section 3.1-3.3 explain the city growth model, Section 3.4 explains projection of urban potentials, and Section 3.5 explains how to projects urban expansion based on the potentials (see, Figure 1). Then, Section 4 explains how to estimate gridded population and gross productivity.

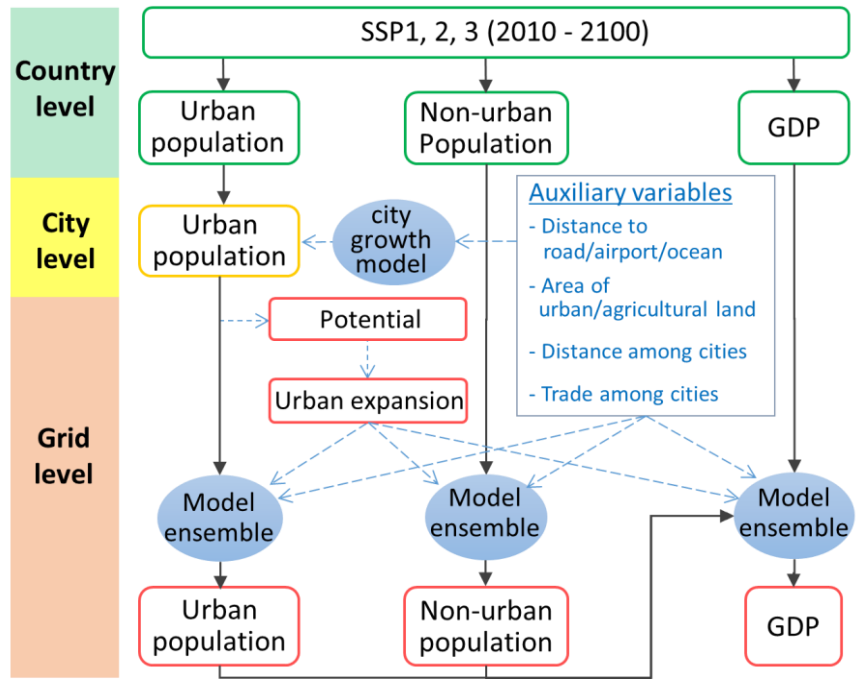

Figure 1: Procedure for population and GDP downscaling. Variables by countries, cities, and grids are coloured by green, yellow, and red, respectively. The black allows represent the downscale procedure while the blue allows represent sub-processing to consider auxiliary variables. As this figure shows, urban population is downscaled from countries to cities to grids, while non-urban population is from countries to grids. GDP is from countries to grids by utilizing downscaled populations.

Table 1: Auxiliary variables.

| Variables | Description | Unit | Source | Year |
|---|---|---|---|---|
| *City pop* | City population | 67,934 cities | SEDAC[1] | 1990, 1995, 2000 |
| *Urban area* | Urban area [km$^2$] | 0.5-degree grids | MODIS[2] | 2000 |
| *Agri area* | Agricultural areas [km$^2$] | | | |
| *Road dens* | Total length [km] of principal roads | | Natural Earth[3] | 2012 |
| *Airport dist* | Distance [km] to the nearest airport | N.A. | | |
| *Ocean dist* | Distance [km] to the nearest ocean | | | 2010 |
| *Trade amount* | Amount of bilateral trade [current US dollars] | Country | CoW[4] | 2009 |

[1] SEDAC (Socioeconomic Data and Applications Center), http://sedac.ciesin.columbia.edu/
[2] MODIS (MODerate resolution Imaging Spectroradiometer), http://modis.gsfc.nasa.gov/
[3] Natural Earth, http://www.naturalearthdata.com/
[4] CoW (The Correlates of War project), http://www.correlatesofwar.org/

## 3. Projection of urban population and urban expansion

### 3.1. City growth model

This study models the population change of 67,934 cities included in the SEDAC dataset (see Table 1) using the following spatial econometric model. The model considers (i) attributes of the cities, (ii) spatial interactions among neighbouring cities, and (iii) global interactions among cities with strong economic connectivity.

$$\Delta \mathbf{p}_{t+5}^{(\log)} = (\rho_{geo}\mathbf{W}_{geo} + \rho_{e1}\mathbf{W}_{e1} + \rho_{e2}\mathbf{W}_{e2})\Delta \mathbf{p}_{t}^{(\log)} + \alpha \mathbf{p}_{t}^{(\log)} + \mathbf{X}_{t}\boldsymbol{\beta} + \boldsymbol{\varepsilon}_{t}, \qquad (1)$$

$$E[\boldsymbol{\varepsilon}_{t}] = \mathbf{0}, \qquad Var[\boldsymbol{\varepsilon}_{t}] = \sigma^2 \mathbf{I}.$$

Suppose that $p_{c,t}$ is the population of city $c$ in year $t$. $\mathbf{p}_t^{(\log)}$ and $\Delta\mathbf{p}_t^{(\log)}$ are $N \times 1$ vectors whose $c$-th elements are $\log(p_{c,t})$ and $\log(p_{c,t}/p_{c,t-5})$, respectively. $\mathbf{X}_t$ is a $N \times K$ matrix of explanatory variables, $\boldsymbol{\varepsilon}_t$ is a $N \times 1$ vector of disturbance with variance $\sigma^2$, $\mathbf{0}$ is a $N \times 1$ vector of zeros, $\mathbf{I}$ is an $N \times N$ identity matrix, $\alpha$ is a coefficient (scalar), and $\boldsymbol{\beta}$ is a $K \times 1$ coefficients vector. Eq.(1) can be derived based on the logistic growth model (e.g., Smith et al., 2002), which is a standard population growth model (see Appendix 1).

Following the literature on spatial econometrics, $\mathbf{W}_{geo}$, $\mathbf{W}^{e1}$, and $\mathbf{W}^{e2}$ are given by row-standardizing [1] (i.e., row sums are scaled to one) $\mathbf{W}^0_{geo}$, $\mathbf{W}^0_{e1}$, and $\mathbf{W}^0_{e2}$, which describe connectivity among cities. $\mathbf{W}^0_{geo}$ is a spatial connectivity matrix whose ($c$, $c'$)-th element is $\exp(-d_{c,c'}/r)$, where $d_{c,c'}$ is the arc distance between cities $c$ and $c'$, and $r$ is a range parameter. For instance, if $r = 100$km, 95% of spill-over effects disappear within 300 km (=3 ×100km; Cressie, 1993). In other words, large $r$ implies global spill-over from cities whereas small $r$ implies local spill-over. $\mathbf{W}^0_{e1}$ and $\mathbf{W}^0_{e2}$ describe economic connectivity. Since we could not find any data on economic connectivity among cities, we approximate it with Eq.(2), which represents estimate of trade amount between cities $c$ and $c'$:

$$\hat{t}_{c,c'} = \frac{p_c}{P_C}\frac{p_{c'}}{P_{C'}}T_{C,C'}, \qquad (2)$$

---
[1] The row standardization makes parameter estimates stable.

where $P_C$ is the population of the country including the $c$-th city, and $T_{C,C'}$ is the amount of trade between countries $C$ and $C'$ (source: CoW data set; see Table 1). Eq.(2) simply distributes the amount of trade, $T_{C,C'}$, in proportion to city populations. The $(c, c')$-th element of $\mathbf{W}^0_{e1}$ is given by $\hat{t}_{c,c'}$ if cities $c$ and $c'$ are in different countries (i.e., $C \neq C'$), and 0 otherwise. By contrast, the $(c, c')$-th elements of $\mathbf{W}^0_{e2}$ are given by $\hat{t}_{c,c'}$ if these cities are in the same country (i.e., $C = C'$), and 0 otherwise. After all, $\mathbf{W}_{e1}$ and $\mathbf{W}_{e2}$ describe international and national economic connectivity, respectively.

If $\rho_{geo}$ is positive, population growth in a city increases the populations in its neighbouring cities. When $\rho_{e1}$ and/or $\rho_{e2}$ is positive, population growth in a city increases the populations in foreign cities with strong economic connectivity. Intuitively speaking, $\rho_{geo}$ and $\rho_{e2}$ capture local interactions, and $\rho_{e1}$ captures global interactions.

*3.2. Parameter estimation*

We use the data of city populations (1990, 1995, 2000) provided by SEDAC (see table 1 and figure 2). In other words, Eq.(1) is estimated while assuming $t = 1995$. The spatial 2-step least squares (2SLS; Kelejian and Prucha, 2002) is used for the estimation[2]. Explanatory variables are road density (Road dens), distance to the nearest airport (Airport dist), and distance to the nearest ocean (Ocean dist; see Table 1), whose coefficients are denoted by $\beta_{road}$, $\beta_{ocean}$, and $\beta_{airport}$, respectively.

Table 2 summarizes the estimated parameters. The table suggests that population increases rapidly in areas with dense road network and good access to airports, although the latter is statistically insignificant. These results are intuitively consistent. The positive sign of $\beta_{ocean}$ suggests that city growth in inland areas tends to be faster than that in bayside cities. This might be because bayside cities are already matured, and their populations are more stable than those of inland cities.

Regarding parameters describing interactions, $\rho^{geo}$ has a statistically significant positive effect, whereas $\rho^{e2}$ does not. Thus, geographic proximity is a significant factor determining local-scale city interactions. On the other hand, $\rho^{e1}$, which quantifies global-scale interactions, is statistically significant. It is suggested that consideration of both local and global-scale interactions is important in city growth modelling.

---

[2] To estimate $r$ in $\mathbf{W}_{geo}$, 2SLS is iterated while varying $r$ values, and the optimal $r$ value, which maximizes the adjusted $R^2$, is identified.

The quasi-adjusted $R^2$ for the population change in 5 years, $\Delta \mathbf{p}_{t+5}$, is 0.401, which is not very accurate. However, the value of $R^2$ for the population after 5 years, $\mathbf{p}_{t+5}$, is 0.998. Since we focus on the latter, the accuracy of the model is sufficient.

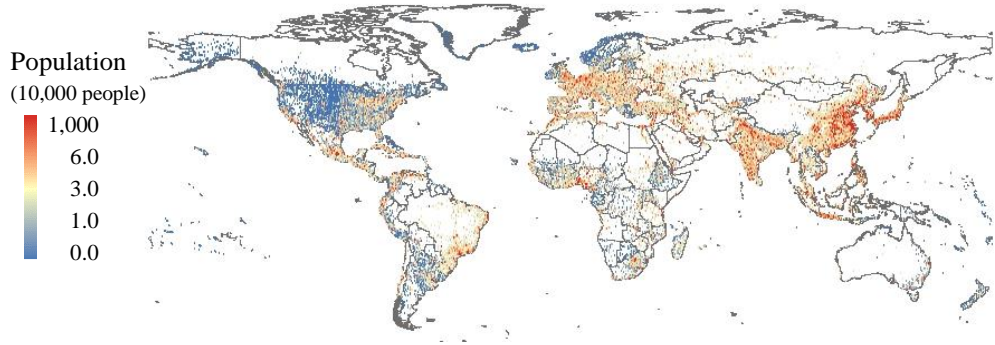

Figure 2: Populations in 67,934 cities (2000).

Source: SEDAC.

Table 2: Parameter estimates.

|  | Estimate | $t$-value |  |
|---|---|---|---|
| Intercept | $-6.19 \times 10^{-4}$ | $-8.12$ | *** |
| $\alpha$ | $1.87 \times 10^{-3}$ | 8.98 | *** |
| $\rho^{geo}$ | $9.56 \times 10^{-1}$ | 188.57 | *** |
| $\rho^{e1}$ | $1.83 \times 10^{-3}$ | 24.95 | *** |
| $\rho^{e2}$ | $4.10 \times 10^{-4}$ | 0.84 |  |
| $\beta_{road}$ | $1.21 \times 10^{-3}$ | 3.46 | *** |
| $\beta_{ocean}$ | $2.10 \times 10^{-4}$ | 2.19 | *** |
| $\beta_{airport}$ | $-1.66 \times 10^{-4}$ | $-0.47$ |  |
| $r$ | 209 |  |  |
| Quasi-adjusted $R^2$ for $\Delta \mathbf{p}_{t+5}$ | 0.405 |  |  |
| Quasi-adjusted $R^2$ for $\mathbf{p}_{t+5}$ | 0.998 |  |  |

[1] * suggests statistical significance at the 1 % level.

*3.3. Projection of city populations*

Since SSP1-3 are globalization, BAU, and fragmentation scenarios, respectively, different levels of international interactions are assumed in each scenario. Specifically, we assume that $\rho^{e1}$ doubles by 2100 in comparison with 2000 in SSP1, $\rho^{e1}$ is constant in SSP2, and $\rho^{e1}$ becomes half the value in 2000 by 2100 in SSP3. In each scenario, the values for $\rho^{e1}$ between 2000 and 2100 are linearly interpolated.

Using the $\rho_{e1}$ values, city populations in 2005, 2010,...2100 are estimated by sequentially applying the city growth model, Eq.(1), which projects the 5-year-after populations.

*3.4. Projection of urbanization potentials*

Increase/decrease of city population encourages/discourages urbanization in the neighbouring areas. Thus, this study evaluates urbanization potential using Eq.(3):

$$q_{g,t}(r') = \sum_c \hat{p}_{c,t} \exp\left(-\frac{d_{c,g}}{r'}\right), \qquad (3)$$

where $\hat{p}_{c,t}$ is the city population in year *t*, which is projected in Section 3.3, $d_{c,g}$ is the arc distance between the *c*-th city and the centre of the *g*-th grid. Eq.(3) increases nearby cities with large population.

Although $r'$ is a range parameter just like $r$ in $\mathbf{W}^0_{geo}$, $r'$ represents the range of spill-over around each city whereas $r$ (= 209 km; see Table 2) represents the range of spill-over across cities. Thus, $r'$ must be smaller than $r$. Considering the consistency with the subsequent urban area projection in Section 3.5, $r'$ is given by a value maximizing the explanatory power of urban potential, $q_{g,t}(r')$, on urban expansion. In other words, $r'$ is estimated by maximizing the adjusted *R*-squares (adj-$R^2$) of the following model:

$$Urban\ Area_{g,2000} = b_0 + q_{g,2000}(r')\,b_q + \varepsilon_{g,2000}, \qquad (4)$$

where $Urban\ Area_{g,2000}$ is the urban area in *g*-th grid in 2000 (source: MODIS; see Table 1), and $\varepsilon_{g,2000}$ denotes disturbance. The estimated parameters in 2000 are $\hat{r}' = 16.4$, $\hat{b}_0 = 21.89$ and $\hat{b}_q = 0.126$. $r' = 16.4$ is assumed for SSP2. On the other hand, $r' = 8.2$ (= 0.5 × 16.4) is assumed for SSP1 to model compact urban growth, while $r' = 32.8$ (= 2.0 × 16.4) is assumed in SSP3 to model dispersed growth.

Figure 3 displays urbanization potentials from Eq.(3) in Europe in 2080. Because of the $r'$ values, potentials in SSP1 are the most compactly distributed while those in SSP3 are the most dispersed. As a result, potentials in SSP3 tend to be high even in between cities where potentials in SSP1 and 2 are small.

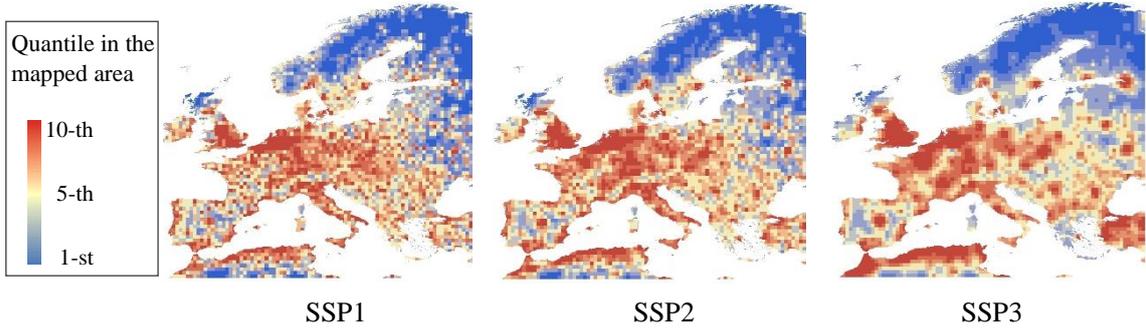
SSP1  SSP2  SSP3
Figure 3: Projected urbanization potential in Europe in 2080.

3.5. *Projection of urban expansion/shrinkage*

    This section projects urban extent based on estimated urbanization potentials (see Figure 1). 5-year change of urban area in each grid is projected by Eq.(5), which is derived from Eq.(4):

$$\Delta Urban\ Area_{g,t+5} = [q_{g,t+5}(r') - q_{g,t}(r')]\hat{b}_q, \quad (5)$$

We also project expansion of non-urban residential areas due to the potentials. This study assumes that non-urban residential areas are proportion to *Agri area* (see Table 1), and the 5-year change is estimated by the following model:

$$\Delta Agri\ Area_{g,t+5} = [q^A_{g,t+5}(r'^A) - q^A_{g,t}(r'^A)]\hat{b}^A_q. \quad (6)$$

Parameters in 2000 are estimated by the adj-$R^2$ maximization of Eq.(4) whose *Urban Area*$_{g,2000}$ is replaced with *Agri Area*$_{g,2000}$. The estimated values are $\hat{r}'^A = 12.1$ and $\hat{b}^A_q = 0.129$. While $b^A_q = 0.129$ is assumed across scenarios, $r'^A$ values in SSP1-3 are given by 6.05, 12.1, and 24.2, respectively, just like $r'$.

    Urban areas and agricultural areas are projected by applying Eqs.(5) and (6) sequentially. In each sequence, if (*Urban Area*$_{g,t+5}$ + *Agri Area*$_{g,t+5}$) exceeds the area of the grid, *Agri area*$_{g,t+5}$ is reduced. *Urban Area*$_{g,2000}$ and *Agri Area*$_{g,2000}$ are used as baseline areas.

    The next section applies the estimated urban and non-urban areas as weights for proportional distribution. In the distribution, the range parameters, $r$, $r'$, and $r'^A$ control share of populations and gross productivity nearby cities. For instance, if $r'$ is very small as in SSP1, most people and gross productivity are concentrated nearby cities. As such, the proportional distribution can describe both urban expansion and shrinkage depending on range parameter values. Similarly, $r'^A$ controls non-urban population distribution. In case of SSP1, the small $r^A$ concentrates non-urban populations into grids with greater *Agri Area* with greater potentials. In other words, non-urban populations are concentrated nearby urban areas. The populations are dispersed in SSP3 whose $r'^A$ value is large.

## 4. Downscaling of SSPs

*4.1. Model*

This section downscales the urban and non-urban populations, and GDPs, respectively, utilizing projected city populations, urbanization potentials, urban areas, and other auxiliary variables summarized in Table 1.

To date, numerous downscale methods have been proposed in quantitative geography, geostatistics, and other fields. The accurateness of the dasymetric mapping, which simply distributes populations in proportion to axillary variables, has been remarked upon in many comparative studies (e.g., Fischer and Langford, 1995; Hawley and Moellering, 2005). We use Eq.(7), which modifies the dasymetric mapping model to consider difference in scenarios[3]:

$$f(a_{g,t,k}) = \frac{\sqrt{\tilde{a}_{g,t}^{ssp} a_{g,t,k}}}{\sum_{g \in C} \sqrt{\tilde{a}_{g,t}^{ssp} a_{g,t,k}}} Y_{C,t}, \qquad (7)$$

where $Y_{C,t}$ is population or GDP in country $C$ including $g$-th grid in year $t$. $\tilde{a}_g^{ssp}$ is a weight to consider urban expansion/shrinkage assumed in each scenario. *Urban area*$_{g,t}$, *Agri area*$_{g,t}$, and *UAgri area*$_{g,t}$ (=*Urban area*$_{g,t}$ +*Agri area*$_{g,t}$), which are projected under each SSP, are used to downscale urban population, non-urban population, and GDP, respectively.

$a_{g,t,k}$ is another weight capturing influence from auxiliary variables, where $k$ is the index of the weights. We are not sure which auxiliary variables are appropriate for $a_{g,t,k}$. Hence, this study uses a weighted average of dasymetric mapping models, which is formulated as follows:

$$\hat{y}_{g,t} = \sum_{k=1}^{K} \omega_{k,t} f(a_{g,t,k}), \qquad (8)$$

where $y_{g,t}$ demotes population/GDP in $g$-th grid, and $\omega_{k,t}$ measures the importance of the $k$-th sub-model, $f(a_{g,t,k})$, in year $t$. The following country level model is obtained by aggregating the grid-level model presented by Eq.(9):

$$Y_{C(g),t} = \sum_{g \in C(g)} \sum_{k=1}^{K} \omega_{k,t} f(a_{g,t,k}). \qquad (9)$$

$\omega_{k,t}$ in the downscale model Eq.(8) is estimated by applying the gradient boosting (Freidman, 2002), which is an ensemble learning technique, for Eq.(9)[4].

---

[3] Square root is used because distribution weights are defined by product of two weight variables.
[4] Result of downscaling depends on which model is used. Ensemble learning decreases uncertainty due to such model selection by taking a weighted average of candidate models just like Eq.(8), while increasing the accuracy of the model by estimating, $\omega_{k,t}$, e.g., by the gradient boosting that sequentially reduces residual errors by fitting small sub-models of Eq.(8) (see, e.g., Bishop, 2006)

The auxiliary variables, $a_{g,t,k}$, are defined by (weight variables) × (control variables). For urban population downscaling, the following weight variables are used: (i) *Urban Area*; (iii) urbanization potential (i.e., $q_{g,t}$); and, (iv) urban population, which is projected by Eq.(1) and aggregated into grids. Because non-urban residents are not in urban area, (i) is replaced with (ii) *Agri Area* to downscale non-urban populations. Regarding GDP, (i + ii) *UAgri area*, (iii), and (iv) are considered. In addition, (v) downscaled urban + non-urban populations are also considered. In other words, gridded gross productivity is estimated after the population downscaling.

Control variables are intended to adjust weighted variables; for example, even if urban areas in two grids are the same, it is reasonable to assign greater weights to the grid with denser road network. Our control variables include (a) Constant, (b) Road dens, (c) Airport dist, and (d) Ocean dist (see Table 1).

As discussed, we define the auxiliary variables, $a_{g,t,k}$, by multiplying the weighted variables and control variables. In other words, we use 12 auxiliary variables, {i, iii, iv} × {a, b, c, d}, for the urban population downscaling, 12 for the non-urban population downscaling ({ii, iii, iv} × {a, b, c, d}), and 16 for the GDP downscaling ({(i+ii), iii, iv, v} × {a, b, c, d}).

*4.2. Result*

Table 3 summarizes the importance of the auxiliary variables, which are estimated by gradient boosting. From this table, the importance of urbanization potential is suggested, especially to explain non-urban populations. Actually, Urban potential explains 55% (SSP1), 54%, (SSP2) and 48% (SSP3) of urban population distributions while 69 %, 68 %, and 64 % of non-urban populations. Regarding urban population downscaling, Distance to the ocean has the biggest contribution (SSP1: 38%, SSP2: 47%, SSP3: 46%). Because many of mega-cities are near the ocean, the result is intuitively reasonable. Concerning non-urban population, distance to principal road has the largest contribution. It is suggested that non-urban population grows along principal roads. The contribution of Road is significant in SSP1 (48%)[5]. It might be because cities are strongly interacted in SSP1, and small cities emerge in between these cities. On the other hand, Ocean is more important than Road in SSP3 (36%).

Distribution of gross productivity, which is estimated by the GDP downscaling, depends on many auxiliary variables. In SSP1, (Urban pop × 1) is estimated the most influential (18%) while (Urban pop × Air) is the secondly influential (14%). Based on the result, city growth and their interaction through airport encourage economic growth in SSP1. By contrast, (Urban potential × Road) and (Urban potential × Air) are significant in SSP3 whose contributions are both about 17%. The result is interpretable that dispersed urbanization yields dispersed economic growth, and the growth is substantial along road network and nearby airports. In short, SSP1 and SSP3 result in compact and dispersed economic growth, respectively, and SSP2 lies in between them.

---

[5] It is calculated by aggregating shares of $a_{g,t,k}$ = (weight variables) × (control variables) whose control variables equal Road (i.e.., 48% = 3% + 3% + 41%).

Figure 4 plots the estimated population distributions in 2080 under SSP1-3. Compared with SSP3, SSP1 and SSP2 show higher population density around mega-cities, including London, Paris, and NY. By contrast, SSP3 has higher and dispersed population density in Africa, West-Middle Asia. Thus, the populations in SSP1 are concentrated while those in SSP3 are dispersed. The concentration and dispersed patterns would be due to the $r'$ and $r'^A$ values given following scenario assumptions. It is verified that the parameters are useful to capture difference in SSPs.

Figure 5 displays the distributions of gross productivity in 2080. Results in SSP1 and SSP2 are relatively similar; both show considerable economic productivity around mega cities (e.g., London and NY). By contrast, economic productivity are small and dispersed in SSP3. Figure 6 displays results of the GDP downscaling in Europe and South-West Asia. In Europe, economic productivity around major cities (e.g., London and Paris) significantly changes depending on SSPs. In South-West Asia, compared with SSP1-2, SSP3 shows lower productivity in urban areas whereas higher productivity in non-urban areas. In other words, SSP3 results in dispersed economic growth. Consider of such difference among SSPs would be important to analyse future climate risks on socioeconomic activities.

Figure 7 compares our population estimates in 2080 in SSP2 with Jones and O'Neill (2016). Estimates of Jones and O'Neill (2016) tend to be overly smoothed (e.g., populations are uniformly distributed in desert areas in Saudi Arabia). It might be because they apply a gravity-based approach, which ignores auxiliary variables. In our result, such over smoothing is not conceivable. It is verified that, while the $r'$ and $r'^A$ parameters are required to consider assumptions in SSPs (i.e., compact/dispersed), consideration of auxiliary variables is also needed to avoid over-smoothing.

Finally, we evaluate accuracy of our downscaling by comparing our population estimates with Gridded Population of the World in 2000 (GPW Version3; source: SEDAC), which is another gridded population database created by aggregating/proportionally distributing administrative data. While resolution of the GPW data depends on country, it is conceivable that data in USA, France, Spain, Portugal, and Japan are high resolution. We apply GPW data in these countries as true population counts. As shown in Figure 8, which compares our estimates and SEDAC estimates in these countries, our estimates is close to the SEDAC estimates. It is verified that our downscale approach is accurate. Of course, the accuracy assessment is only for 2000; it would be an important research topic to evaluate accuracy and uncertainty of our downscaling result in the future.

Table 3: Estimated importance of auxiliary variables in 2080 ($a_{g,k}$ = offset variables × control variables).

| Offset variables | Urban area | | | | Urban pop | | | | Urban potential | | | |
|---|---|---|---|---|---|---|---|---|---|---|---|---|
| Control variables | 1 | Road | Air | Ocean | 1 | Road | Air | Ocean | 1 | Road | Air | Ocean |
| Urban population SSP1 | 0.02 | 0.10 | 0.07 | 0.11 | 0.02 | 0.01 | 0.03 | 0.11 | 0.05 | 0.19 | 0.15 | 0.16 |
| Urban population SSP2 | 0.09 | 0.05 | 0.05 | 0.10 | 0.02 | 0.02 | 0.03 | 0.11 | 0.03 | 0.13 | 0.12 | 0.26 |
| Urban population SSP3 | 0.07 | 0.03 | 0.05 | 0.10 | 0.08 | 0.07 | 0.06 | 0.08 | 0.03 | 0.04 | 0.13 | 0.28 |

| Offset variables | Agri area | | | | Urban pop | | | | Urban potential | | | |
|---|---|---|---|---|---|---|---|---|---|---|---|---|
| Control variables | 1 | Road | Air | Ocean | 1 | Road | Air | Ocean | 1 | Road | Air | Ocean |
| Non-urban population SSP1 | 0.03 | 0.04 | 0.07 | 0.03 | 0.02 | 0.03 | 0.03 | 0.06 | 0.08 | 0.41 | 0.13 | 0.07 |
| Non-urban population SSP2 | 0.04 | 0.03 | 0.09 | 0.03 | 0.02 | 0.01 | 0.03 | 0.07 | 0.08 | 0.37 | 0.12 | 0.11 |
| Non-urban population SSP3 | 0.07 | 0.02 | 0.10 | 0.04 | 0.01 | 0.01 | 0.03 | 0.09 | 0.05 | 0.17 | 0.19 | 0.23 |

| Offset variables | Urban + Agri area | | | | Urban pop | | | | Urban potential | | | | SSP pop | | | |
|---|---|---|---|---|---|---|---|---|---|---|---|---|---|---|---|---|
| Control variables | 1 | Road | Air | Ocean | 1 | Road | Air | Ocean | 1 | Road | Air | Ocean | 1 | Road | Air | Ocean |
| GDP SSP1 | 0.07 | 0.01 | 0.04 | 0.05 | 0.18 | 0.10 | 0.14 | 0.04 | 0.06 | 0.03 | 0.03 | 0.04 | 0.02 | 0.06 | 0.09 | 0.04 |
| GDP SSP2 | 0.10 | 0.02 | 0.03 | 0.05 | 0.14 | 0.09 | 0.08 | 0.03 | 0.08 | 0.08 | 0.05 | 0.06 | 0.08 | 0.03 | 0.07 | 0.03 |
| GDP SSP3 | 0.01 | 0.05 | 0.01 | 0.05 | 0.10 | 0.09 | 0.01 | 0.05 | 0.09 | 0.17 | 0.17 | 0.01 | 0.09 | 0.02 | 0.07 | 0.01 |

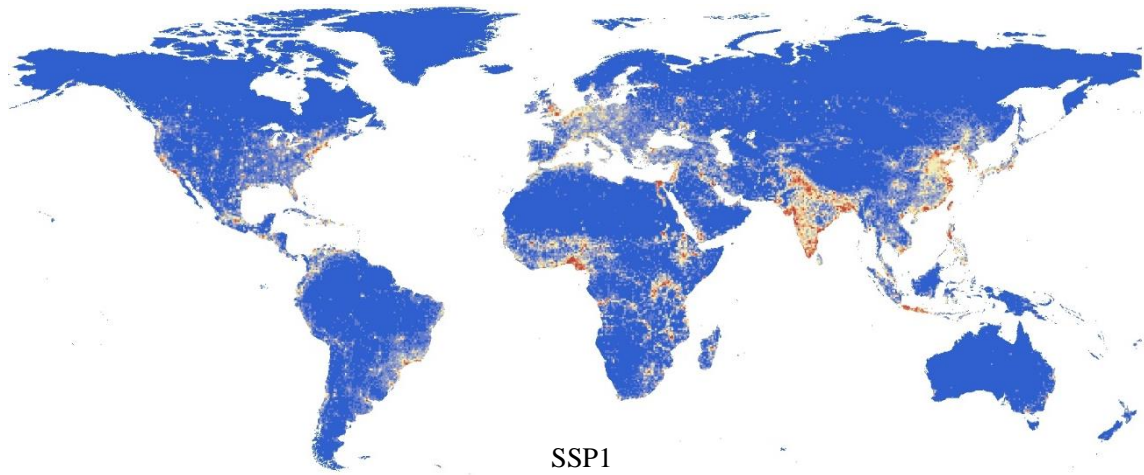
SSP1

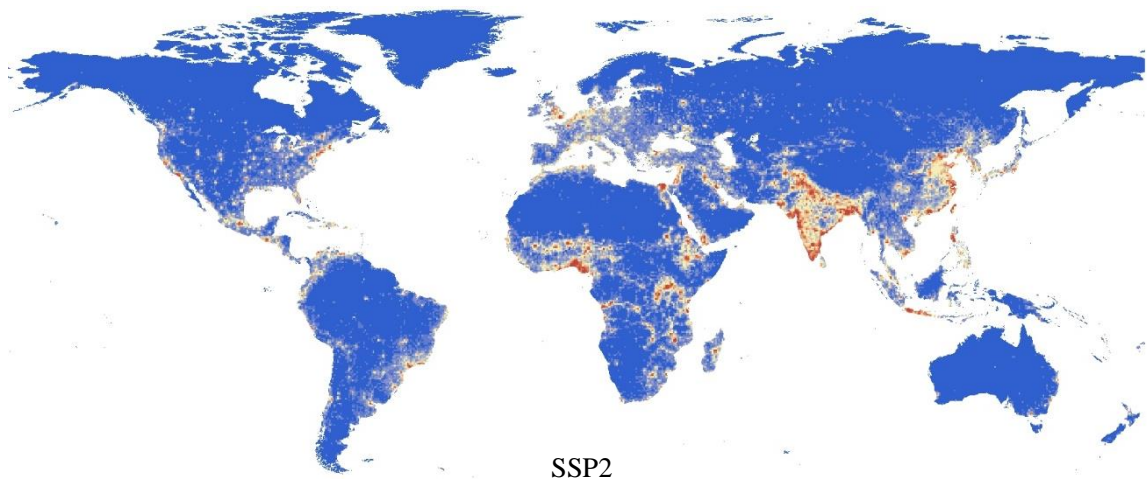
SSP2

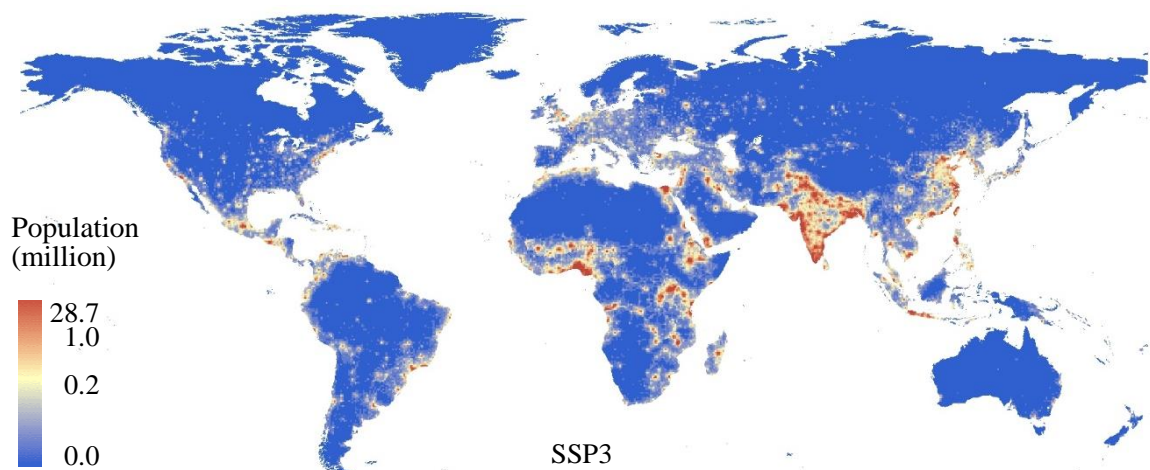
Population (million)

28.7
1.0
0.2
0.0

SSP3

Figure 4: Downscaled population distributions in 2080.

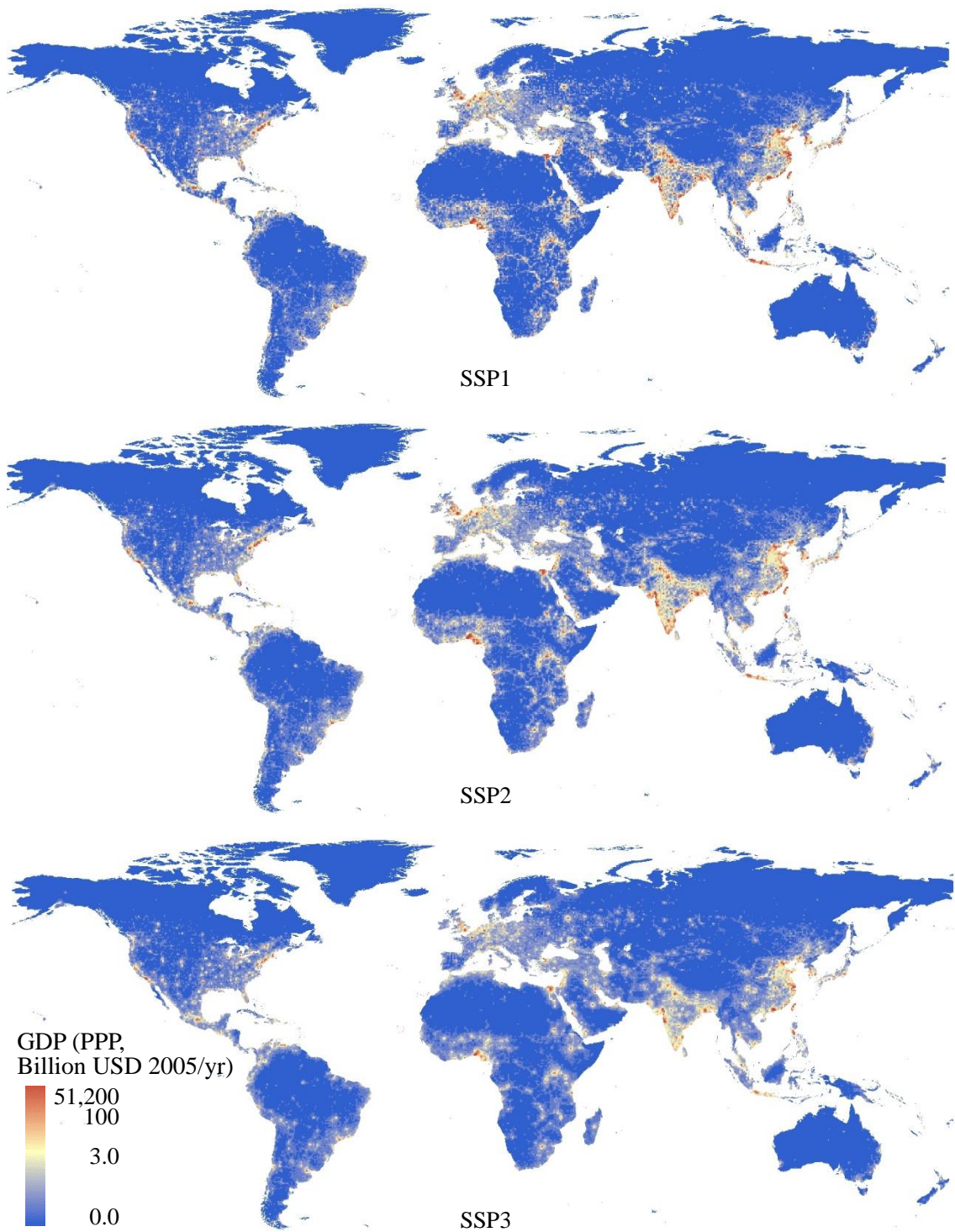

Figure 5: Downscaled gross productivities in 2080.

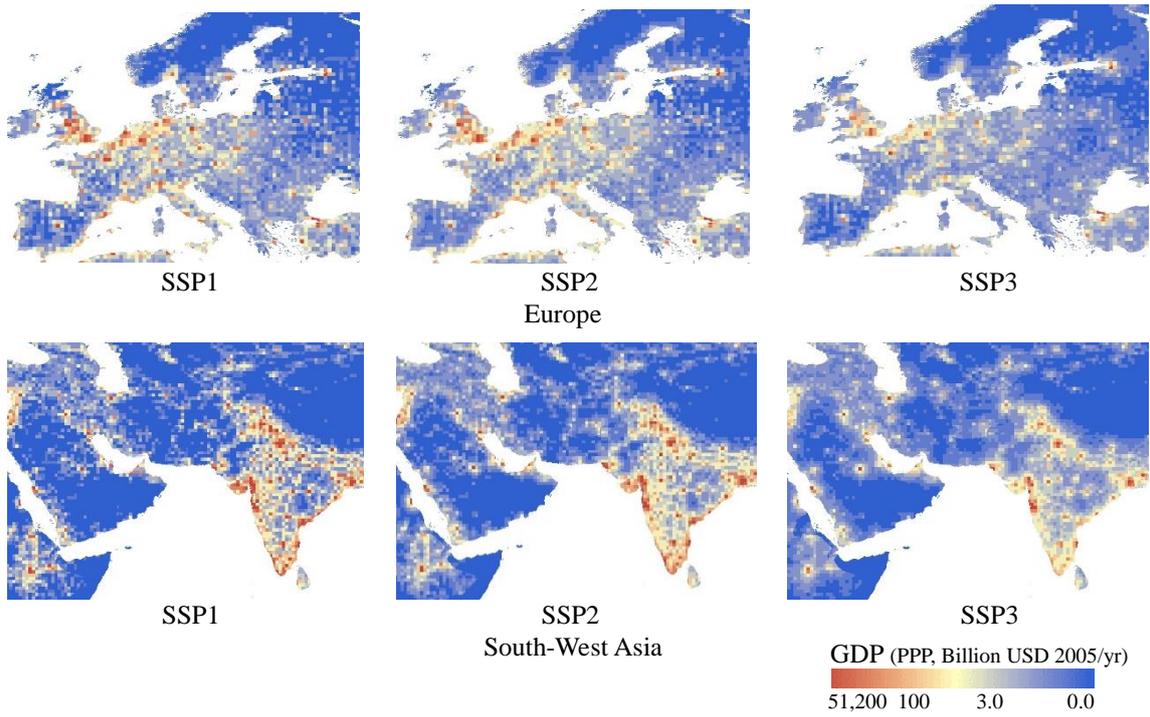

Figure 6: Downscaled gross productivities in 2080 (South and West Asia, and Europe).

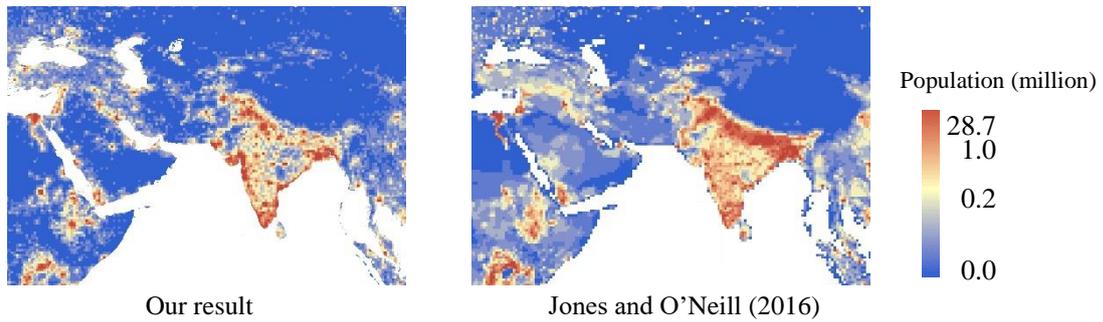

Figure 7: Comparison of estimated populations in South-West Asia in 2080 (SSP2)

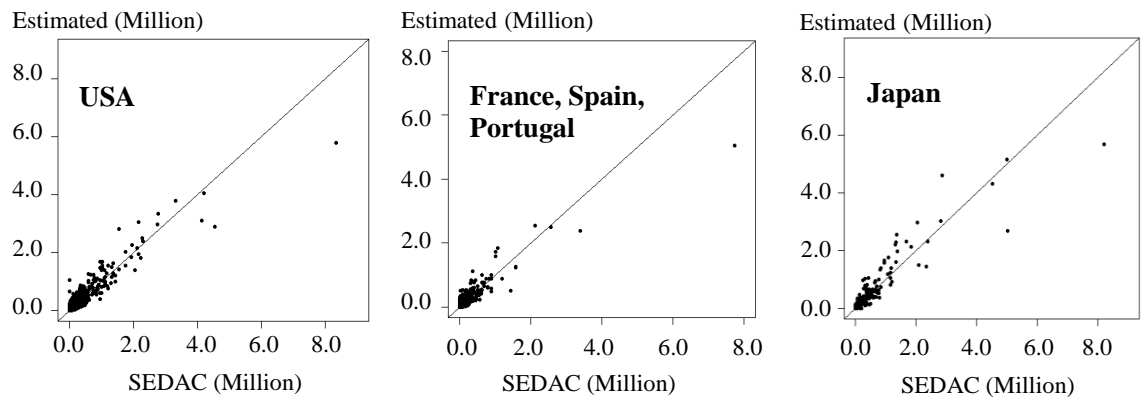

Figure 8: Comparison of our population grid and SEDAC population grid in 2000.

## 5. Concluding remarks

This study downscales SSP scenarios into 0.5-degree grids, using a model to consider spatial and economic interaction among cities and an ensemble learning technique to utilize multiple auxiliary variables accurately. The downscaling result suggests that SSP1, which refers to the sustainable scenario, yields a compact population distribution relative to SSP3, which denotes the fragmentation scenario. The results also show that GDP growth in major metropolitan areas changes significantly depending on the scenarios. These results are intuitively consistent. The consideration of such differences is critical to the estimation of grid level $CO_2$ emissions, disaster risks, energy demand, and other variables determining future sustainability and resiliency.

Nonetheless, various other important issues require further study. First, a spatially finer auxiliary dataset is needed to increase the accuracy of the downscaling. For example, additional city-level data, such as industrial structure, detailed road network, and traffic volume, are required to describe urban phenomena such as industrial agglomeration, growth of the transportation network, and the birth of new cities. Since using these factors can increase the uncertainty of downscaling, it is crucial to employ a robust estimation approach, such as ensemble learning (applied in this paper) or Bayesian estimation (as done by Raftery et al. (2012) for population projection).

Second, downscaling to finer grids is required. Although 0.5-degree grids are sufficient to evaluate socioeconomic activities in each region, these grids are not sufficient to quantify urban form, such as compact and disperse. Finer grids, such as 1 km grids, are required to analyze impact of urban form on climate change mitigation and adaption. High-resolution auxiliary variables would be needed to achieve it.

Third, it is needed to validate our estimates based on not only current data but also historical data. Unavailability of historical data of road network, airport locations, GDP, and so on, make the validation difficult. Development of these historical dataset would be beneficial for both validation and advancement of projection/downscale approaches.

Fourth, it is important to discuss how we may utilize our estimates for city-level policy making. The project titled World Urban Database and Access Portal Tools (WUDAPT: http://www.wudapt.org/) is an interesting activity in this respect. The project aims to (i) collect data describing urban forms and functions (e.g., land cover, building structure, and building allocations), (ii) utilize the data to classify urban areas into 17 Local Climate Zones (LCZs; Stewart and Oke, 2012), and (iii) design universal policies for each of the LCZs toward improving climate resilience. While LCZs classify urban areas based on their influence on the ambient local climate, distributions of population and gross productivity are key factors determining $CO_2$ emissions and amount of wasted heat. Thus, our downscaled populations and GDPs might help design LCZs and devise appropriate policies.

Our downscaling results are available from "Global dataset of gridded population and GDP scenarios," which is provided by the Global Carbon Project, National Institute of Environmental Studies (http://www.cger.nies.go.jp/gcp/population-and-gdp.html). This dataset summarizes population and GDP scenarios in 0.5 × 0.5 degree grids between 1980 and 2100 by 10 years. The gridded data in 2020 - 2100 are estimated by downscaling country level SSP1-3 scenarios (SSP database: https://secure.iiasa.ac.at/web-apps/ene/SspDb/dsd?Action=htmlpage&page=about) as explained in this manuscript, where those in 1980 - 2010 are estimated by applying the same downscale method to actual populations and GDPs by country (source: IMF data; http://www.imf.org/data).

**Acknowledgement**

This study was funded by the Global Climate Risk Management Strategies (S10) Project of the Ministry of the Environment, Japan. We also acknowledge the generous support from Dr. Hajime Seya (Kobe University, Japan).

**Appendix 1**: Relationship between the logistic growth model and the spatial econometric model

The logistic growth model, which is a popular population growth model, is formulated as follows:

$$\Delta p_{c,t+5} = r p_{c,t} \left(1 - \frac{p_{c,t}}{M}\right), \qquad (A.1)$$

where $p_{c,t}$ is the population of city $c$ in year $t$, $\Delta p_{c,t+5} = p_{c,t+5} / p_{c,t}$, $M$ is the carrying capacity in a city, and $r$ is a parameter. A preliminary analysis suggests that the larger cities in our city dataset, which tend to have greater population growth (this is also conceivable from the positive value of $\alpha$ in table 2) and carrying capacity, $M$, do not have any negative influence on population growth. Thus, we assume that $M$ is sufficiently large, and $\frac{p_{c,t}}{M} \approx 0$. In other words, Eq.(A.1) is $\Delta p_{c,t+5} = r p_{c,t}$, which is also known as the exponential growth model. The exponential growth model can be further expanded to consider auxiliary variables, using a Cobb–Douglas-type expression, as follows:

$$\Delta p_{c,t+5} = r p_{c,t}^{\alpha} \left(\prod_{k}^{K-1} z_{k,c,t}^{\beta_k}\right) e_{c,t}, \qquad (A.2)$$

where $\alpha$ and $\beta_k$ are parameters, $z_{k,c,t}$ is the $k$-th auxiliary variable in $c$-th city in year $y$, and $e_{c,t}$ is a positive disturbance. Both $z_{k,c,t}$ and $e_{c,t}$ must be positive. The log-transformation of Eq.(A.2) yields

$$\log(\Delta p_{c,t+5}) = \log(r) + \alpha \log(p_{c,t}) + \sum_{k=1}^{K-1} \log(z_{k,c,t})\beta_k + \varepsilon_{c,t}. \quad (A.3)$$

$$E[\varepsilon_{c,t}] = 0, \qquad Var[\varepsilon_{c,t}] = \sigma^2,$$

where $\beta_0 = \log(r)$ and $\varepsilon_{c,t} = \log(e_{c,t})$. Here, it is assumed that $\varepsilon_{c,t}$ is independent and identically distributed (i.i.d.). Then, Eq.(A.3) takes the following matrix expression:

$$\Delta \mathbf{p}_{t+5}^{(\log)} = \alpha \mathbf{p}_t^{(\log)} + \mathbf{X}_t \boldsymbol{\beta} + \boldsymbol{\varepsilon}_t. \quad (A.4)$$

$$E[\boldsymbol{\varepsilon}_t] = \mathbf{0}, \qquad Var[\boldsymbol{\varepsilon}_t] = \sigma^2 \mathbf{I},$$

where $\boldsymbol{\beta} = [\beta_0, \beta_1,..., \beta_{K-1}]'$, $\mathbf{X}_t$ is a matrix whose first column is a vector of ones, and the elements in the other $K$-1 columns are are given by $\log(z_{k,c,t})$.

The spatial econometric model Eq.(1) is obtained by introducing the spatial and economic interaction effects into Eq.(A.4).